\documentclass[conference]{IEEEtran}
\IEEEoverridecommandlockouts

\usepackage{cite}
\usepackage{amsmath,amssymb,amsfonts}
\usepackage{algorithmic}
\usepackage{graphicx}
\usepackage{textcomp}
\usepackage{braket}
\usepackage{xcolor}
\usepackage{url}
\usepackage{tabularx}
\usepackage{array}
\usepackage{longtable}
\usepackage{booktabs}
\def\BibTeX{{\rm B\kern-.05em{\sc i\kern-.025em b}\kern-.08em
    T\kern-.1667em\lower.7ex\hbox{E}\kern-.125emX}}
\begin{document}

\title{RL-Guided Quantum-ALNS for Constrained VRP\\
{\large \textbf{IEEE International Conference on Quantum Computing and Engineering (IEEE Quantum Week), September 13–18, 2026, Toronto, ON, Canada}}
\thanks{This research is funded by the Canada Research Program in Disruptive Transportation Systems and Services (CRC-2021-0048) and the Canada First Research Excellence program. We are thankful to PINQ$^2$ and IBM for the QPU access.}
}

\author{\IEEEauthorblockN{Farzan Moosavi}
\IEEEauthorblockA{\textit{Laboratory of Innovations in Transportation} \\
\textit{Toronto Metropolitan University}\\
Toronto, Canada \\
0009-0008-8490-3851}
\and
\IEEEauthorblockN{Bilal Farooq}
\IEEEauthorblockA{\textit{Laboratory of Innovations in Transportation} \\
\textit{Toronto Metropolitan University}\\
Toronto, Canada \\
0000-0003-1980-5645}
}

\maketitle

\begin{abstract}

Abstract—This study develops a hybrid quantum--classical
framework for constrained vehicle routing problems, focusing on
the pickup-and-delivery problem with time windows. Instead of
casting the full routing problem as a stand-alone quantum
optimization task, we embed shallow quantum samplers inside the
repair phase of an Adaptive Large Neighbourhood Search (ALNS)
heuristic. A Deep Q-Network controller decides whether each
reduced repair subproblem should be handled by a classical repair
heuristic or by a quantum sampler, using features that describe
the local repair structure and predicted hardware reliability. IBM
Heron experiments are used to calibrate an empirical noise-aware
model for local quantum repair circuits. Across the tested
instances, quantum repair is admissible in only about 16\% of
reduced repair states and is not superior on average. However,
under selected matched repair budgets, quantum-enabled repair
reduces the final gap relative to standard ALNS in 29 of 36 tested
settings. These results suggest that near-term quantum sampling is
most useful as a selective local repair mechanism rather than as a
replacement for classical routing heuristics.


\end{abstract}

\begin{IEEEkeywords}
Vehicle Routing Problem, Quantum Computation, Adaptive Large Neighborhood Search, Quantum Circuit Design.
\end{IEEEkeywords}

\section{Introduction}

Quantum computing has attracted growing interest in transportation and logistics as a possible computational paradigm for hard combinatorial optimization problems, including routing and delivery applications \cite{b1,b2,b3}. One important example is the pickup-and-delivery problem with time windows (PDPTW), in which vehicles must satisfy paired pickup and delivery requests under capacity and time-window constraints. The PDPTW is NP-hard and becomes increasingly challenging as operational constraints scale, frequently causing even strong classical heuristics to struggle to efficiently explore large neighborhoods.

Early quantum work in routing focused mainly on quantum annealing \cite{b4}. More recently, gate-based quantum computing has gained attention for combinatorial search through methods such as the Quantum Approximate Optimization Algorithm (QAOA) \cite{b5}, Grover-style search \cite{b6}, and learning-based quantum circuit design \cite{b7}. Despite this promise, solving highly constrained routing problems end-to-end on gate-based hardware remains prohibitive in the Noisy Intermediate-Scale Quantum (NISQ) era. Standard Quadratic Unconstrained Binary Optimization (QUBO) formulations require many penalty terms to enforce routing constraints, which can distort the energy landscape and reduce the probability of sampling feasible solutions \cite{b8}. Alternative encodings, such as colored-permutation formulations \cite{b9} and constrained mixers \cite{b10}, attempt to restrict the search to feasible regions, but they typically require deeper circuits and additional entangling operations \cite{b11,b12}. As a result, the practical usefulness of a quantum formulation is limited by three interacting NISQ factors: penalty-induced landscape distortion, depth limitations on superconducting hardware \cite{b13,b14}, and the difficulty of classical variational parameter tuning, which is often sensitive to noise, initialization, and instance structure \cite{b15,b16,b17, b18}.

Because purely quantum approaches to constrained vehicle routing remain unviable at realistic problem scales \cite{b19,b20}, a more practical near-term direction is to embed quantum routines within classical optimization frameworks and invoke quantum sampling only on smaller subproblems \cite{b21,b22,b23}. This hybrid strategy has shown promise in related settings, including warm-started circuit initialization \cite{b24}, quantum-enhanced Markov chain Monte Carlo (MCMC) sampling \cite{b25}, column-generation frameworks \cite{b26}, and clustered routing decompositions \cite{b27,b28}. Motivated by these developments, we study a hybrid framework for the PDPTW that embeds gate-based quantum sampling within an Adaptive Large Neighborhood Search (ALNS) procedure. ALNS is a natural host for such integration because its destroy--repair mechanism repeatedly creates reduced repair subproblems \cite{b29}.

However, invoking a quantum processor at every ALNS repair step is impractical. Cloud-based quantum execution incurs substantial latency due to queueing, calibration, and communication overhead. In addition, NISQ noise can dominate the circuit output when reduced repair neighborhoods are too small or too simple to benefit from quantum sampling. In such cases, the decoded quantum repairs can yield solutions inferior to fast classical insertions based on local cost information. This motivates a learned control policy that invokes quantum repair only when the expected benefit justifies the estimated hardware cost and reliability risk \cite{b30,b31, b32}.

To address this issue, we propose a dynamic repair framework in which each classical ALNS destroy phase defines a reduced repair subproblem, and a Deep Q-Network (DQN) controller selects one admissible repair action from a mixed classical--quantum action set. The controller observes reduced-repair features, including structural entropy, feasible-candidate ratios, search history, and noise-aware indicators. The quantum branch uses QAOA and EfficientSU2 circuit families, while the classical branch retains standard heuristic repair operators as discussed in \cite{b29}. In addition, an empirical IBM-backend benchmark model predicts the reliability and cost of candidate quantum calls. These predictions are used as masking conditions, allowing the DQN to avoid quantum execution when the predicted noise level, qubit requirement, or execution burden exceeds acceptable thresholds.

The main methodological contribution of this work is a noise-aware reinforcement-learning policy for dynamically orchestrating classical and quantum repair operators within an ALNS backbone. Conceived as a practical NISQ-era hybrid method, the proposed framework does not claim unconditional quantum advantage. Rather, its goal is to identify exactly when quantum-guided repair has practical heuristic value in constrained routing environments, and when classical repair should remain dominant.

\section{Methodology}

\begin{figure*}[htbp]
\centering
\includegraphics[width=0.8\textwidth]{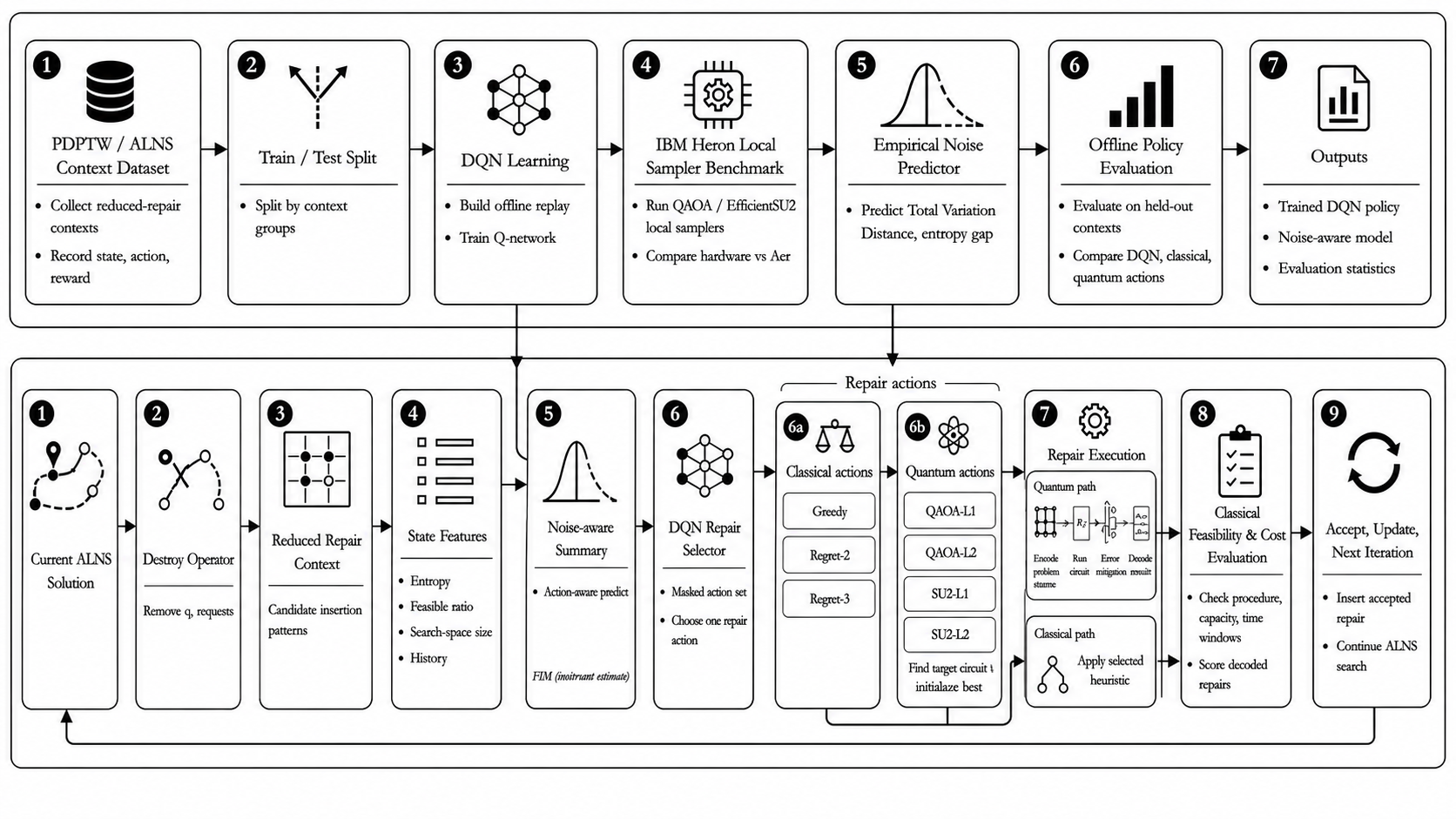}
\caption{Overview of the proposed DQN-guided quantum--classical ALNS framework.
The upper panel shows offline data collection, DQN training, IBM Heron sampler
benchmarking, empirical calibration of the noise predictor, and held-out policy
evaluation. The lower panel shows the online ALNS rollout, where the learned
controller selects between classical and quantum repair actions using structural
repair-context features and noise-aware reliability-cost estimates.}
\label{fig:architecture}
\end{figure*}

The proposed method embeds a learned quantum--classical repair selector inside
an Adaptive Large Neighbourhood Search (ALNS) framework for the pickup and
delivery problem with time windows (PDPTW). The framework has two phases, as
shown in Fig.~\ref{fig:architecture}. In the offline phase, reduced repair
contexts are collected from ALNS rollouts, local quantum repair circuits are
benchmarked on IBM Heron hardware, an empirical noise-aware predictor is
calibrated, and a Deep Q-Network (DQN) repair selector is trained. In the online
phase, the trained DQN selects, at each ALNS iteration, whether the destroyed
requests should be repaired by a classical heuristic or by a quantum sampler.
All candidate repairs are finally evaluated using the original PDPTW feasibility
constraints and objective before ALNS acceptance.

\subsection{Problem Setting and ALNS Backbone}

We consider the PDPTW on a directed graph \(G=(V,A)\), where
\(V=\{0\}\cup P\cup D\). Node \(0\) is the depot, \(P\) is the set of pickup
nodes, and \(D\) is the set of delivery nodes. Each request
\(r\in\mathcal{R}\) consists of a pickup--delivery pair \((p_r,d_r)\). Each
node \(i\in V\) has service duration \(\sigma_i\), time window
\([e_i,\ell_i]\), and load demand \(q_i\), with
\(q_{p_r}>0\) and \(q_{d_r}=-q_{p_r}\). A fixed fleet
\(K=\{1,\ldots,m\}\) of homogeneous vehicles with capacity \(Q\) is used.

A solution \(S=\{R_k\}_{k\in K}\) is a set of vehicle routes, where
\[
R_k=(v_{k,0},v_{k,1},\ldots,v_{k,n_k},v_{k,n_k+1}),
\qquad
v_{k,0}=v_{k,n_k+1}=0 .
\]
The objective is to minimize total travel cost:
\begin{equation}
\min f(S)=
\sum_{k\in K}\sum_{h=0}^{n_k}
c_{v_{k,h},v_{k,h+1}} .
\label{eq:objective}
\end{equation}
A feasible solution must serve every request exactly once, assign each pickup
and delivery to the same vehicle, visit every pickup before its delivery,
respect vehicle capacity, and satisfy all time windows. If \(B_{k,h}\) is the
service start time and \(L_{k,h}\) is the vehicle load after servicing
\(v_{k,h}\), then feasibility requires
\[
0\leq L_{k,h}\leq Q,
\qquad
e_{v_{k,h}}\leq B_{k,h}\leq \ell_{v_{k,h}}.,
\]
together with temporal consistency along each route.

ALNS iteratively applies a destroy operator followed by a repair operator. At
iteration \(t\), a destroy operator removes \(q_t\) requests from the current
solution \(S_t\), producing a removed-request set
\(\mathcal{R}_t^{-}\) and a partial solution \(\bar{S}_t\). A repair action
then reinserts the removed requests and produces a candidate solution \(S'_t\).
Feasible candidates are accepted according to the simulated-annealing rule
\begin{equation}
P_{\mathrm{acc}}(S'_t)=
\begin{cases}
1, & f(S'_t)\leq f(S_t),\\[1mm]
\exp\!\left(-\dfrac{f(S'_t)-f(S_t)}{T_t}\right),
& f(S'_t)>f(S_t).
\end{cases}
\label{eq:sa_accept}
\end{equation}
The proposed method keeps the destroy phase and the ALNS acceptance rule
unchanged. Learning is introduced only in the repair-selection stage.

\subsection{Reduced Repair Model and Quantum Sampling}

After destruction, the local repair task is restricted to reinserting
\(\mathcal{R}_t^{-}\) into \(\bar{S}_t\). For each removed request
\(r\in\mathcal{R}_t^{-}\), let \(\mathcal{A}_{r,t}\) be the set of candidate
insertion patterns. A pattern \(a=(k,\alpha,\beta)\) inserts the pickup and
delivery of request \(r\) into route \(k\) at positions \(\alpha\) and
\(\beta\), with \(\beta>\alpha\), so pickup--delivery precedence is enforced at
the pattern level.

The incremental cost of assigning pattern \(a\) to request \(r\) is
\begin{equation}
\Delta_{r,a}=f(\bar{S}_t\oplus a)-f(\bar{S}_t),
\label{eq:delta_ra}
\end{equation}
where \(\bar{S}_t\oplus a\) denotes the partial solution after applying the
pattern. Pairwise interaction effects between two insertion choices are
approximated by
\begin{equation}
\Gamma_{r,a;s,b}
=
f(\bar{S}_t\oplus a\oplus b)
-f(\bar{S}_t)-\Delta_{r,a}-\Delta_{s,b}.
\label{eq:gamma_pair}
\end{equation}
If two patterns cannot be selected together, the pair is included in the
conflict set \(\mathcal{C}_t\) and penalized through
\[
\widetilde{\Gamma}_{r,a;s,b}
=
\Gamma_{r,a;s,b}
+
P_{\mathrm{pair}}\mathbb{I}\{((r,a),(s,b))\in\mathcal{C}_t\}.
\]

The reduced repair model is then written as a binary quadratic objective. Let
\(x_{r,a}=1\) if pattern \(a\in\mathcal{A}_{r,t}\) is selected for request
\(r\), and \(x_{r,a}=0\) otherwise. The exactly-one-pattern condition is
enforced using a one-hot penalty. The binary reduced repair objective is
\begin{align}
E_t^{\mathrm{bin}}(x)
=&
\sum_{r\in\mathcal{R}_t^{-}}
\sum_{a\in\mathcal{A}_{r,t}}
\Delta_{r,a}x_{r,a}
\nonumber\\
&+
\sum_{\substack{r,s\in\mathcal{R}_t^{-}\\ r<s}}
\sum_{a\in\mathcal{A}_{r,t}}
\sum_{b\in\mathcal{A}_{s,t}}
\widetilde{\Gamma}_{r,a;s,b}x_{r,a}x_{s,b}
\nonumber\\
&+
P_{\mathrm{one}}
\sum_{r\in\mathcal{R}_t^{-}}
\left(1-\sum_{a\in\mathcal{A}_{r,t}}x_{r,a}\right)^2 .
\label{eq:binary_energy}
\end{align}
This reduced objective is not the full PDPTW objective; it is a local
approximation used to generate candidate repairs. Every decoded repair is still
checked against the original PDPTW constraints.

For quantum repair actions, \eqref{eq:binary_energy} is mapped to an Ising cost
Hamiltonian using \(x_i=(1-z_i)/2\), yielding
\begin{equation}
H_C^{(t)}
=
\sum_i h_i^{(t)}Z_i
+
\sum_{i<j}J_{ij}^{(t)}Z_iZ_j
+
c^{(t)}I .
\label{eq:cost_hamiltonian}
\end{equation}
The implementation considers two circuit families: QAOA and hardware-efficient
SU(2) circuits. The actions QAOA-L1 and QAOA-L2 correspond to one- and
two-layer QAOA circuits, while SU2-L1 and SU2-L2 correspond to one- and
two-layer EfficientSU2 circuits. Circuit parameters are not optimized online.
Instead, a small bank of fixed and random initializations is probed with a
small shot budget, and the initialization whose measured output entropy is
closest to the target entropy of the selected action is used for final sampling.
Measured bitstrings are decoded into insertion-pattern selections; invalid
bitstrings that violate the one-hot condition are discarded.

\subsection{Noise-Aware DQN Repair Controller}

The DQN acts as a high-level repair selector. It does not choose insertion
patterns directly. Instead, after the destroy step has produced the reduced
repair context, it selects one repair action \(u_t\) from the mixed action set
\begin{equation}
\mathcal{U}=\mathcal{U}_C\cup\mathcal{U}_Q,
\end{equation}
where
\begin{equation}
\mathcal{U}_C=
\{\mathrm{greedy},\mathrm{regret\mbox{-}2},\mathrm{regret\mbox{-}3}\},
\end{equation}
and
\begin{equation}
\mathcal{U}_Q=
\{\mathrm{QAOA\mbox{-}L1},\mathrm{QAOA\mbox{-}L2},
\mathrm{SU2\mbox{-}L1},\mathrm{SU2\mbox{-}L2}\}.
\end{equation}

At iteration \(t\), the DQN observes a state vector
\begin{equation}
s_t=[e_t,z_t,h_t],
\label{eq:state_blocks}
\end{equation}
where \(e_t\) contains entropy descriptors of the reduced repair landscape,
\(z_t\) contains reduced-neighbourhood statistics, and \(h_t\) contains
search-history features. All continuous features are scaled using statistics
computed from the training data before being passed to the DQN.

The entropy block is
\begin{equation}
e_t=
[
H_t^{\mathrm{ins}},
H_t^{\mathrm{conf}},
H_t^{\mathrm{slack}},
H_t^{\mathrm{load}},
H_t^{\mathrm{struct}}
].
\label{eq:entropy_block}
\end{equation}
These entropy quantities are normalized Shannon-type descriptors. They are used
only as state features and are not added to the reduced repair objective or to
the reward.

For each removed request \(r\in\mathcal{R}_t^{-}\), insertion costs are first
converted into a probability vector over candidate patterns using a
temperature-scaled softmax:
\begin{equation}
\pi_{r,a,t}
=
\frac{
\exp[-(\Delta_{r,a}-\Delta_{r,t}^{\min})/\tau_{\Delta}]
}{
\sum_{b\in\mathcal{A}_{r,t}}
\exp[-(\Delta_{r,b}-\Delta_{r,t}^{\min})/\tau_{\Delta}]
},
\label{eq:entropy_softmax_compact}
\end{equation}
where
\begin{equation}
\Delta_{r,t}^{\min}
=
\min_{b\in\mathcal{A}_{r,t}}\Delta_{r,b},
\qquad
\tau_{\Delta}>0 .
\end{equation}
The normalized insertion entropy of request \(r\) is
\begin{equation}
H^{\mathrm{ins}}_{r,t}
=
-\frac{
\sum_{a\in\mathcal{A}_{r,t}}\pi_{r,a,t}\log \pi_{r,a,t}
}{
\log |\mathcal{A}_{r,t}|
},
\label{eq:entropy_insert_compact}
\end{equation}
when \(|\mathcal{A}_{r,t}|>1\), and is set to zero otherwise. The
context-level insertion entropy is the average
\begin{equation}
H_t^{\mathrm{ins}}
=
\frac{1}{|\mathcal{R}_t^{-}|}
\sum_{r\in\mathcal{R}_t^{-}}H^{\mathrm{ins}}_{r,t}.
\label{eq:context_insert_entropy_compact}
\end{equation}
Low insertion entropy indicates that a few insertion patterns dominate, whereas
high insertion entropy indicates that several patterns have comparable costs.

The conflict density is computed from the incompatible pattern-pair set
\(\mathcal{C}_t\):
\begin{equation}
\rho_t^{\mathrm{conf}}
=
\frac{|\mathcal{C}_t|}
{
\sum_{\substack{r,s\in\mathcal{R}_t^{-}\\r<s}}
|\mathcal{A}_{r,t}|\,|\mathcal{A}_{s,t}|
},
\label{eq:conflict_density_compact}
\end{equation}
with \(\rho_t^{\mathrm{conf}}=0\) if the denominator is zero. Its binary entropy
is denoted by \(H_t^{\mathrm{conf}}\). The slack and load entropy terms,
\(H_t^{\mathrm{slack}}\) and \(H_t^{\mathrm{load}}\), are computed by
discretizing feasible insertion slack values and residual capacity margins into
fixed histograms whose bin ranges are determined from the training data. Empty
histograms are assigned entropy zero. The aggregate structural entropy is
\begin{equation}
H_t^{\mathrm{struct}}
=
\frac{
H_t^{\mathrm{ins}}
+
H_t^{\mathrm{conf}}
+
H_t^{\mathrm{slack}}
+
H_t^{\mathrm{load}}
}{4}.
\label{eq:struct_entropy_compact}
\end{equation}

The reduced-neighbourhood block is
\begin{equation}
z_t=
[
\rho_t^{\mathrm{feas}},
\rho_t^{\mathrm{conf}},
\ell_t,
q_t,
\xi_t,
\nu_t^{\min},
\bar{\ell}_t,
1
],
\label{eq:neighbourhood_block}
\end{equation}
where \(\rho_t^{\mathrm{feas}}\) is the ratio of individually feasible insertion
patterns, \(\ell_t\) is the normalized log-size of the reduced candidate space
\(\Omega_t\), \(q_t=|\mathcal{R}_t^{-}|\) is the destroy size, \(\xi_t\) is the
normalized stagnation count, \(\nu_t^{\min}\) is the noise-aware reliability
feature defined below, \(\bar{\ell}_t=\log(1+|\Omega_t|)\), and the final entry
is a constant bias feature.

The search-history block is
\begin{equation}
h_t=
[
\eta_t,
I_{t-1}^{\mathrm{acc}},
I_{t-1}^{\mathrm{best}},
\vartheta_t
],
\label{eq:history_block}
\end{equation}
where \(\eta_t\) is the normalized iteration index,
\(I_{t-1}^{\mathrm{acc}}\) indicates whether the previous candidate was
accepted, \(I_{t-1}^{\mathrm{best}}\) indicates whether it improved the
incumbent best solution, and \(\vartheta_t=T_t/T_0\) is the normalized
simulated-annealing temperature.

The noise-aware feature is obtained from an empirical predictor calibrated on
paired noiseless and IBM-backend benchmark runs of the local quantum repair
circuits. For each candidate quantum action \(u\in\mathcal{U}_Q\), the
predictor takes a circuit-context vector \(c_t(u)\), including the circuit
family, layer count, shot budget, destroy size, number of candidate patterns,
number of qubits, transpiled depth, two-qubit gate count, and transpiler
optimization level. It outputs
\begin{equation}
g_{\psi}(c_t(u))
=
\bigl(
\widehat{D}^{\mathrm{TV}}_t(u),
\widehat{\Delta H}_t(u),
\widehat{L}^{\mathrm{feas}}_t(u),
\widehat{L}^{\mathrm{best}}_t(u),
\widehat{\tau}^{\mathrm{hw}}_t(u)
\bigr),
\label{eq:noise_predictor_compact}
\end{equation}
where the components estimate total variation distance from the noiseless
distribution, entropy change, feasible-sampling loss, best-solution sampling
loss, and hardware latency. For two output distributions \(P\) and \(Q\) over
bitstrings \(y\), the total variation distance is
\begin{equation}
TVD(P,Q)
=
\frac{1}{2}\sum_y |P(y)-Q(y)|.
\label{eq:tvd_compact}
\end{equation}

The predicted quantities are scaled to \([0,1]\) and combined into an
action-specific reliability-cost score \(\nu_t(u)\). Larger values indicate
less reliable or more costly quantum execution. The DQN state uses the
context-level summary
\begin{equation}
\nu_t^{\min}
=
\min_{u\in\mathcal{U}_Q}\nu_t(u),
\label{eq:noise_min_feature}
\end{equation}
with \(\nu_t^{\min}=1\) if no quantum action is available. This feature
summarizes the best predicted quantum option in the current repair context.

Not every action is admissible in every state. Classical repair actions are
admissible whenever a repair attempt can be constructed from the current reduced
context. A quantum action \(u\in\mathcal{U}_Q\) is admissible only if
\begin{equation}
|\Omega_t|\leq \Omega_u^{\max},
\qquad
\nu_t(u)\leq \nu^{\max}.
\label{eq:quantum_mask}
\end{equation}
Thus, the empirical predictor affects the controller in two ways: the
context-level value \(\nu_t^{\min}\) is used as a state feature, while the
action-specific score \(\nu_t(u)\) is used to mask quantum actions that are too
large or predicted to be too unreliable.

\subsection{Training and Online Execution}

Training data are collected as replay transitions
\begin{equation}
(s_t,u_t,r_t,s_{t+1},m_t,m_{t+1},d_t),
\label{eq:transition_tuple}
\end{equation}
where \(m_t\) and \(m_{t+1}\) are the current and next action masks, and
\(d_t\) is a terminal indicator. The reward is computed after the selected
repair action generates a candidate and after the candidate is checked for
PDPTW feasibility. Let \(S_t^{\star}\) denote the incumbent best solution before
evaluating \(S'_t\), and let
\(I_t^{\mathrm{q}}=\mathbb{I}\{u_t\in\mathcal{U}_Q\}\). The rollout reward is
\begin{equation}
r_t=
\begin{cases}
12.5-0.1I_t^{\mathrm{q}},
& S'_t \text{ feasible and } f(S'_t)<f(S_t^{\star}),\\
6.5-0.1I_t^{\mathrm{q}},
& S'_t \text{ feasible and } f(S'_t)<f(S_t),\\
2.5-0.1I_t^{\mathrm{q}},
& S'_t \text{ feasible and accepted},\\
-0.1I_t^{\mathrm{q}},
& \text{otherwise}.
\end{cases}
\label{eq:dqn_reward}
\end{equation}

The cases are evaluated sequentially. Thus, incumbent-improving repairs receive
the largest reward, accepted non-improving repairs receive a smaller positive
reward, and quantum actions incur a small penalty to discourage unnecessary
quantum calls.

The DQN is trained with a Double-DQN target. Let \(Q_{\theta}\) be the online
network and \(Q_{\theta^-}\) the target network. With discount factor
\(\gamma\), the target is
\begin{equation}
y_t=
r_t+
(1-d_t)\gamma
Q_{\theta^-}(s_{t+1},u_{t+1}^{\star}),
\label{eq:ddqn_target}
\end{equation}
where
\[
u_{t+1}^{\star}
=
\arg\max_{u\in\mathcal{U}(s_{t+1})}
Q_{\theta}(s_{t+1},u).
\]
Masked actions are excluded from the maximization. The network is trained by
minimizing
\begin{equation}
\mathcal{L}(\theta)=
\mathbb{E}
\left[
\left(
y_t-Q_{\theta}(s_t,u_t)
\right)^2
\right].
\label{eq:dqn_loss}
\end{equation}

During online execution, the trained DQN observes the reduced repair context,
selects an admissible repair action, and returns a repaired candidate solution
to the ALNS loop. Classical actions apply their corresponding repair heuristic.
Quantum actions sample the reduced binary repair model, decode valid bitstrings
into candidate repairs, and return the best feasible candidate under the
original PDPTW objective. The final acceptance decision is always made by the
unchanged ALNS acceptance rule in \eqref{eq:sa_accept}.

\begin{table}[!t]
\caption{Main notation used in the methodology.}
\label{tab:comp_notation}
\centering
\scriptsize
\renewcommand{\arraystretch}{1.08}
\setlength{\tabcolsep}{2.5pt}
\begin{tabularx}{\columnwidth}{p{0.22\columnwidth}p{0.25\columnwidth}X}
\toprule
\textbf{Category} & \textbf{Symbol} & \textbf{Description} \\
\midrule

\textbf{Problem}
& \(G=(V,A)\) & Directed graph with node set \(V\) and arc set \(A\) \\
& \(\mathcal{R}\) & Set of pickup--delivery requests \\
& \((p_r,d_r)\) & Pickup and delivery nodes of request \(r\) \\
& \(K\) & Set of homogeneous vehicles \\
& \(Q\) & Vehicle capacity \\
& \(f(S)\) & Total travel cost of solution \(S\) \\
& \(B_{k,h}\) & Service start time at position \(h\) of route \(k\) \\
& \(L_{k,h}\) & Vehicle load after servicing position \(h\) of route \(k\) \\

\midrule

\textbf{ALNS}
& \(S_t\) & Current solution at iteration \(t\) \\
& \(S'_t\) & Candidate solution after repair \\
& \(S_t^{\star}\) & Incumbent best solution before evaluating \(S'_t\) \\
& \(\bar{S}_t\) & Partial solution after destruction \\
& \(\mathcal{R}^{-}_t\) & Removed-request set \\
& \(q_t\) & Destroy size, \(q_t=|\mathcal{R}^{-}_t|\) \\
& \(T_t\) & Simulated-annealing temperature \\

\midrule

\textbf{Repair}
& \(\mathcal{A}_{r,t}\) & Candidate insertion-pattern set for request \(r\) \\
& \(a=(k,\alpha,\beta)\) & Insertion pattern on vehicle \(k\) \\
& \(\Omega_t\) & Reduced candidate-combination space \\
& \(\Delta_{r,a}\) & Incremental cost of inserting \(r\) using pattern \(a\) \\
& \(\Gamma_{r,a;s,b}\) & Pairwise interaction between two insertion choices \\
& \(\mathcal{C}_t\) & Set of incompatible pattern pairs \\
& \(P_{\mathrm{pair}}\) & Penalty for incompatible pattern pairs \\
& \(P_{\mathrm{one}}\) & One-hot selection penalty \\
& \(E_t^{\mathrm{bin}}(x)\) & Binary reduced repair objective \\

\midrule

\textbf{Entropy}
& \(N_t^{\mathrm{pat}}\) & Number of candidate insertion patterns \\
& \(\rho_t^{\mathrm{conf}}\) & Conflict density \\
& \(H_t^{\mathrm{ins}}\) & Context-level insertion entropy \\
& \(H_t^{\mathrm{conf}}\) & Conflict-density entropy \\
& \(H_t^{\mathrm{slack}}\) & Time-window slack entropy \\
& \(H_t^{\mathrm{load}}\) & Residual-capacity entropy \\
& \(H_t^{\mathrm{struct}}\) & Aggregate structural entropy score \\
& \(\rho_t^{\mathrm{feas}}\) & Feasible-pattern ratio \\

\midrule

\textbf{Quantum}
& \(x_{r,a}\) & Binary variable for pattern \(a\) of request \(r\) \\
& \(n_t\) & Number of qubits / binary variables \\
& \(H_C^{(t)}\) & Cost Hamiltonian for iteration \(t\) \\
& \(Z_i,X_i\) & Pauli-\(Z\) and Pauli-\(X\) operators \\
& \(p,L\) & QAOA depth and EfficientSU2 layer count \\
& \(\boldsymbol{\phi}_u\) & Circuit parameters for quantum action \(u\) \\
& \(\bar{H}(u)\) & Target output entropy for quantum action \(u\) \\

\midrule

\textbf{Noise / DQN}
& \(\nu_t(u)\) & Action-specific reliability-cost score \\
& \(\nu_t^{\min}\) & Best available quantum reliability-cost feature \\
& \(s_t\) & DQN state representation \\
& \(u_t\) & Selected repair action \\
& \(m_t\) & Action mask \\
& \(I_t^{\mathrm{q}}\) & Quantum-action indicator \\
& \(Q_{\theta},Q_{\theta^-}\) & Online and target Q-networks \\
& \(y_t\) & Double-DQN target value \\
& \(\mathcal{L}(\theta)\) & DQN training loss \\
\bottomrule
\end{tabularx}
\end{table}

\section{Results and Analysis}

This section evaluates the proposed DQN-guided quantum--classical ALNS
framework from three perspectives. First, we analyze the offline reduced-repair
dataset to characterize when quantum repair actions are admissible and how
their sampled candidates compare with classical repair actions. Second, we
evaluate online ALNS behavior through convergence curves and final-gap
distributions. Third, we compare fixed-budget repair policies under different
PDPTW parameter settings. All algorithmic evaluations are performed in
simulation using instances derived from the Li and Lim PDPTW benchmark
\cite{bLL,bLLsintef}.

Throughout the results, the term \emph{DQN-guided hybrid ALNS} refers to the
proposed controller that dynamically selects repair actions from the mixed
classical--quantum action set \(\mathcal{U}=\mathcal{U}_C\cup\mathcal{U}_Q\).
When reporting fixed-budget comparisons, quantum-enabled repair variants are
reported separately to isolate the effect of the quantum sampler under matched
repair budgets. Lower objective values, relative gaps, and elapsed times
indicate better performance.

\subsection{Experimental Setup}

The offline dataset contains \(120{,}000\) reduced repair contexts collected
from ALNS rollouts. The online ALNS experiments evaluate five random seeds on
small instances with \(|\mathcal{R}|\in\{15,20\}\) and larger instances with
\(|\mathcal{R}|\in\{35,50\}\). Unless otherwise stated, the online experiments
use destroy sizes \(q_t\in\{2,3,4,5\}\), candidate caps in
\(\{2,3,4\}\), \(K=6\) vehicles, \(\mathrm{tw\_tightness}=0.5\), and
\(\mathrm{capacity\_slack}=0.3\).

The parameters \(\mathrm{tw\_tightness}\) and
\(\mathrm{capacity\_slack}\) control different aspects of instance difficulty.
Lower \(\mathrm{tw\_tightness}\) corresponds to looser time windows, whereas
lower \(\mathrm{capacity\_slack}\) corresponds to tighter vehicle-capacity
conditions. Therefore, the fixed-budget results are interpreted directly by
their parameter pair
\((\mathrm{tw\_tightness},\mathrm{capacity\_slack})\), rather than by a single
global difficulty label.

The fixed-budget experiments scale to
\(|\mathcal{R}|\in\{100,150,200\}\) and evaluate shot budgets
\(\{16,128,1024\}\) under two parameter settings:
\[
(\mathrm{tw\_tightness},\mathrm{capacity\_slack})=(0.15,0.15)
\]
and
\[
(\mathrm{tw\_tightness},\mathrm{capacity\_slack})=(0.85,0.85).
\]
The DQN is trained using the Double-DQN target in
\eqref{eq:ddqn_target}. The empirical noise-aware module is calibrated using an
IBM-backend benchmark dataset obtained from local quantum repair samplers, while
the full ALNS rollout experiments are executed in simulation.

\begin{table}[!t]
\caption{Hardware validation for local quantum repair samplers.}
\label{tab:hardware_aer_compact}
\centering
\footnotesize
\begin{tabular}{lccc}
\toprule
\textbf{Metric} & \textbf{Hardware} & \textbf{Aer} & \textbf{\(p\)-value} \\
\midrule
Feasible probability        & 0.093 & 0.090 & 0.006 \\
Best-solution probability   & 0.016 & 0.016 & 0.233 \\
\midrule
Overall TVD mean / median   & 0.389 / 0.307 & -- & -- \\
16-shot TVD mean / median   & 0.580 / 0.688 & -- & -- \\
1024-shot TVD mean / median & 0.209 / 0.161 & -- & -- \\
\bottomrule
\end{tabular}

\vspace{1mm}
\begin{minipage}{0.95\linewidth}
\footnotesize
\textit{Notes:} \(p\)-values are computed using two-sided Wilcoxon signed-rank
tests on matched hardware--Aer runs. TVD denotes total variation distance
between empirical hardware and Aer output distributions.
\end{minipage}
\end{table}

Table~\ref{tab:hardware_aer_compact} summarizes the hardware validation results
for the local quantum repair samplers. The feasible-solution probability is
slightly higher on hardware than in Aer, \(0.093\) versus \(0.090\), and the
matched Wilcoxon test indicates that this small difference is statistically
detectable (\(p=0.006\)). In contrast, the best-solution sampling probability is
nearly identical across hardware and Aer, and the difference is not statistically
significant (\(p=0.233\)).

The empirical output distributions nevertheless differ substantially, with an
overall mean and median TVD of \(0.389\) and \(0.307\), respectively. The estimated distributional discrepancy is larger at low shot counts and smaller
at high shot counts, decreasing from a mean/median TVD of \(0.580/0.688\) at
16 shots to \(0.209/0.161\) at 1024 shots.
These observations motivate the use of an empirical noise-aware predictor in the rollout controller, rather than assuming that noiseless simulation and hardware sampling are interchangeable.

\subsection{Offline Action-Space Analysis}

To better understand the decision environment faced by the DQN controller, we
analyze the offline dataset of \(120{,}000\) reduced PDPTW repair contexts. The
main goal is to characterize when quantum actions are admissible and how their
best achievable rewards compare with those of classical actions in the same
state.

\begin{figure*}[!ht]
\centering
\includegraphics[width=0.8\textwidth]{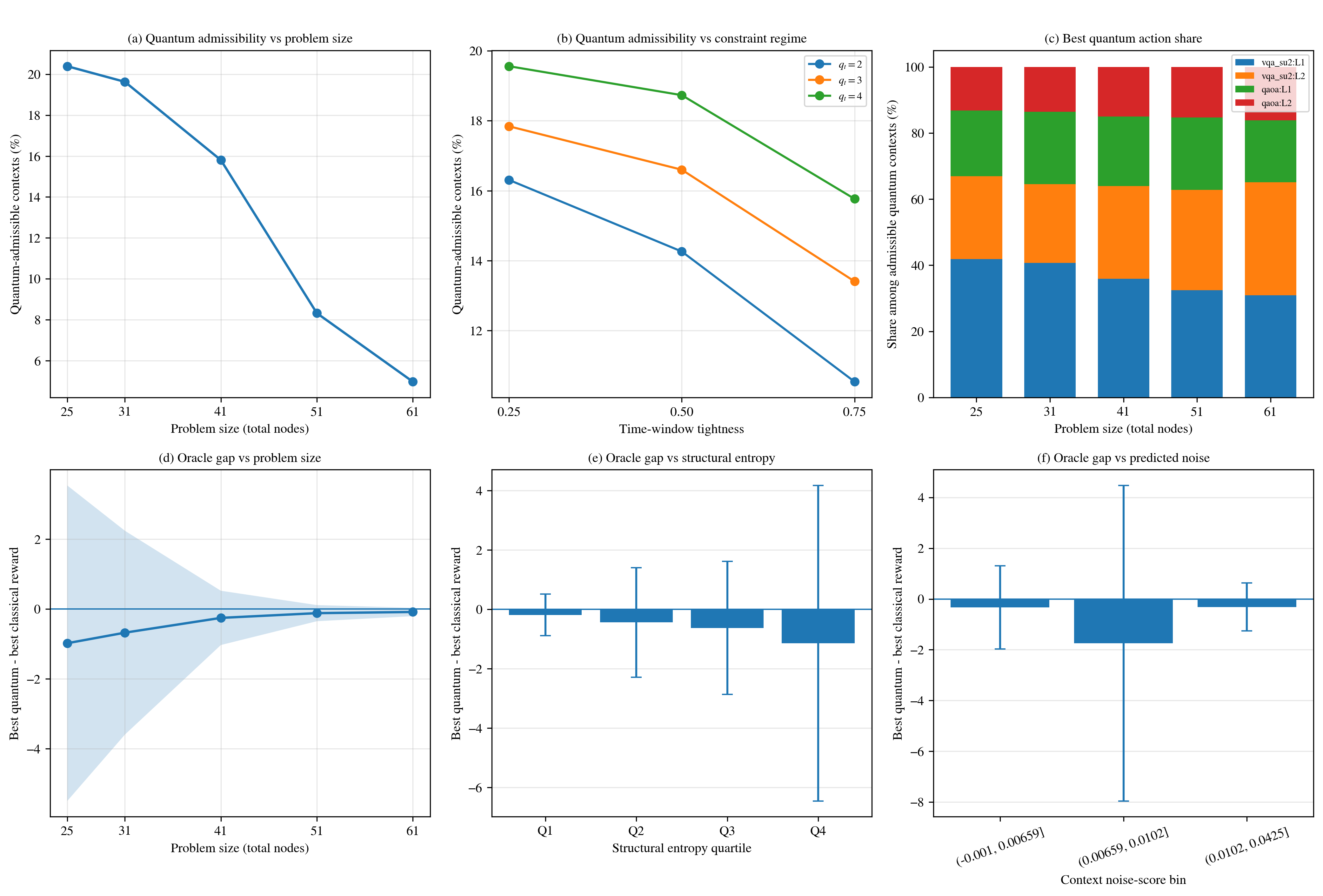}
\caption{Offline action-space analysis over the reduced-repair dataset.
(a) Fraction of contexts with at least one admissible quantum action versus
problem size. (b) Quantum-admissible fraction across time-window regimes and
destroy sizes \(q_t\). (c) Share of the strongest admissible quantum action
among quantum-admissible contexts. (d) Oracle reward gap between the best
admissible quantum action and the best admissible classical action versus
problem size. (e) Oracle reward gap versus structural-entropy quartile.
(f) Oracle reward gap versus predicted reliability-cost-score bin. Error bars
in (d)--(f) denote one standard deviation.}
\label{fig:offline_usage}
\end{figure*}

As shown in Fig.~\ref{fig:offline_usage}(a), the hybrid action space is highly
state-dependent. At least one admissible quantum action is available in only
15.96\% of all reduced repair contexts. Moreover, quantum admissibility
declines as the problem size increases, dropping from roughly 20\% at 25 nodes
to about 5\% at 61 nodes. This confirms that the controller operates in a
predominantly classical regime, with quantum repair available only in a limited
subset of states.

Fig.~\ref{fig:offline_usage}(b) shows how admissibility varies with
\(\mathrm{tw\_tightness}\) and destroy size \(q_t\). Under the instance
generation used in this study, lower \(\mathrm{tw\_tightness}\) corresponds to
looser time windows. The figure therefore indicates that quantum actions are
more often admissible in temporally looser contexts and when more requests are
removed. This trend is consistent with the masking mechanism introduced in the
methodology, since larger or more constrained reduced repair spaces are less
likely to satisfy the admissibility conditions for quantum actions. Among
contexts in which at least one quantum action is admissible, shallow
hardware-efficient circuits are selected more often as the strongest quantum
option: EfficientSU2 accounts for about 65.0\% of the best admissible quantum
actions, while QAOA accounts for the remaining 35.0\%, as shown in
Fig.~\ref{fig:offline_usage}(c).

To compare the strongest admissible quantum and classical options within the
same reduced state, we define the oracle reward gap
\begin{equation}
\Delta_t^{\mathrm{qc}}
=
\max_{u\in\mathcal{U}_Q(s_t)} r(s_t,u)
-
\max_{u\in\mathcal{U}_C(s_t)} r(s_t,u),
\label{eq:oracle_gap}
\end{equation}
where \(\mathcal{U}_Q(s_t)=\mathcal{U}(s_t)\cap\mathcal{U}_Q\) and
\(\mathcal{U}_C(s_t)=\mathcal{U}(s_t)\cap\mathcal{U}_C\). This quantity is
evaluated only for contexts in which \(\mathcal{U}_Q(s_t)\neq\emptyset\). A
positive value indicates that the best admissible quantum action attains a
higher reward than the best admissible classical action in that state.

Over the admissible-quantum subset of the dataset, the mean oracle reward gap
is \(-0.5968\), indicating that classical repair remains stronger on average
under the current noisy simulation and IBM-backend benchmark model. This
pattern is also visible in Fig.~\ref{fig:offline_usage}(d)--(f). The mean
oracle gap remains negative across all problem sizes, although it becomes
closer to zero for larger instances. It also remains negative across all
structural-entropy quartiles and all predicted noise-score bins. The lowest
mean value appears in the highest structural-entropy quartile, suggesting that
highly ambiguous repair contexts remain challenging for the current quantum
repair strategies. Likewise, the middle noise bin shows the lowest mean oracle
gap among the noise-score groups.

At the same time, the variability in Fig.~\ref{fig:offline_usage}(d)--(f) is
substantial, especially in higher-entropy and intermediate-noise regimes. The
wide error bars indicate marked heterogeneity across reduced repair contexts.
Thus, although quantum actions are not superior on average, they can still be
competitive in selected states. This observation supports the main motivation
for the proposed controller: quantum repair should not be treated as a blanket
replacement for classical heuristics, but rather as a conditional capability
that is invoked only when the reduced repair context appears favorable.

Overall, the offline analysis does not support a claim of average quantum
superiority. Instead, it supports the design of a state-dependent repair
controller. The role of the DQN is precisely to identify those contexts in
which quantum repair is admissible and potentially useful, based on structural
descriptors such as \(H_t^{\mathrm{struct}}\), reduced-space size, and the
noise-aware reliability feature.

\subsection{Online ALNS Performance}

We evaluate online search behavior by embedding the learned repair controller
inside the ALNS loop and comparing it with three baselines: Regret-2 repair,
standard Classical ALNS, and Bandit-guided ALNS. Performance is measured using
the relative incumbent gap to the best final objective obtained across all
methods for the same instance--seed pair, referred to as the \emph{seed-best}
objective. Thus, lower values indicate that a method is closer to the best
solution observed under matched experimental conditions.

\subsubsection{Convergence Profile}

\begin{figure*}[!h]
\centering
\includegraphics[width=0.8\textwidth]{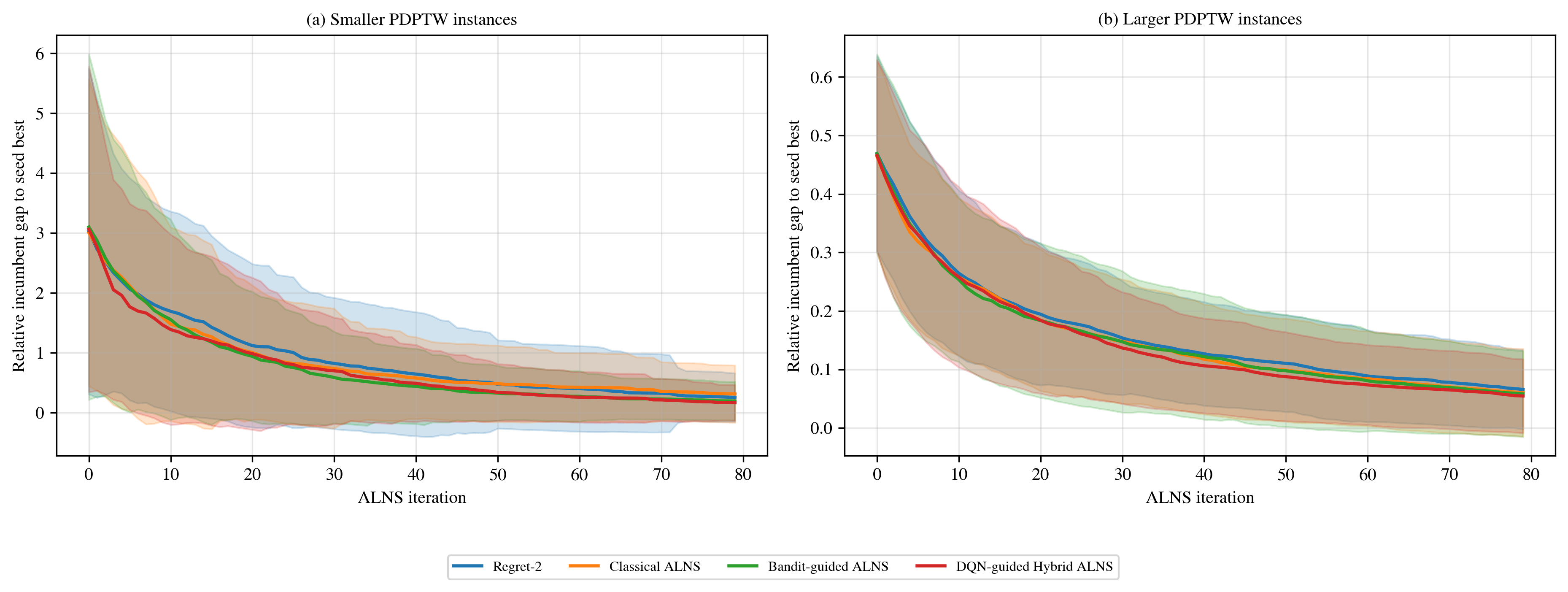}
\caption{Convergence profiles of four ALNS variants on two PDPTW scales. The
vertical axis reports the relative incumbent gap to the seed-best final
objective; lower values are better. Solid lines show the mean across matched
runs, and shaded bands denote one standard deviation. (a) Smaller PDPTW
instances. (b) Larger PDPTW instances.}
\label{fig:convergence_subplots}
\end{figure*}

Fig.~\ref{fig:convergence_subplots} shows the evolution of the relative
incumbent gap over ALNS iterations. Across both instance groups, all methods
improve rapidly in the early phase and then gradually plateau, which is
consistent with standard destroy--repair search behavior.

For the smaller instances in Fig.~\ref{fig:convergence_subplots}(a),
DQN-guided Hybrid ALNS achieves the lowest mean end-of-run gap. It reaches an
average final gap of approximately \(0.20\), compared with about \(0.23\) for
Bandit-guided ALNS and above \(0.30\) for Regret-2. Classical ALNS and Regret-2
converge to higher final gaps in this setting. This indicates that the learned
controller improves the mean search trajectory on the smaller instances.

For the larger instances in Fig.~\ref{fig:convergence_subplots}(b), the
differences between methods are smaller. DQN-guided Hybrid ALNS still obtains
the lowest mean end-of-run gap, approximately \(0.055\), but Bandit-guided ALNS
and Classical ALNS remain close. Therefore, the convergence advantage of the
DQN-guided controller is visible in both groups, but it is more pronounced on
the smaller instances.

The shaded regions show one standard deviation across matched runs and indicate
substantial variability, especially in the early and middle stages of the
search. The convergence results should therefore be interpreted as evidence of
improved average behavior rather than uniform dominance in every individual run.

\subsubsection{Final Gap Versus Number of Requests}

\begin{figure*}[!h]
\centering
\includegraphics[width=0.8\textwidth]{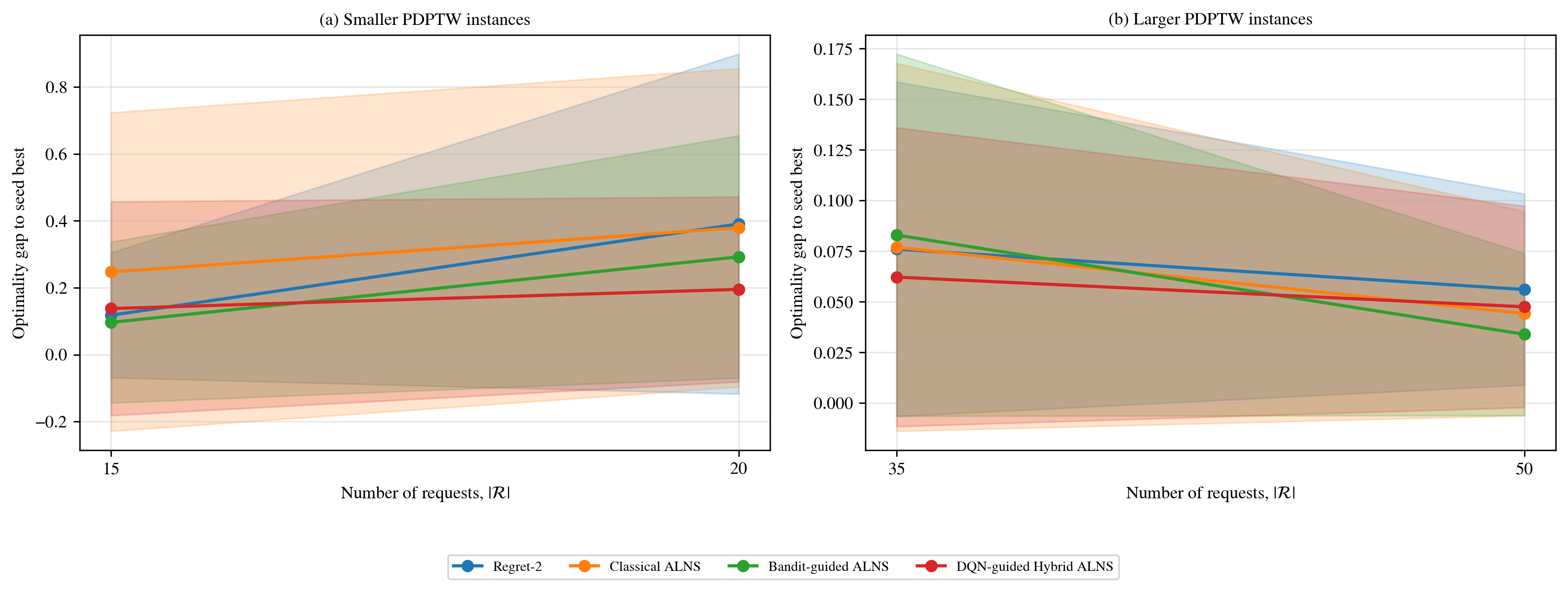}
\caption{Final relative gap to the seed-best solution versus the number of
requests \(|\mathcal{R}|\). Lower values are better. Solid lines show the mean
across matched runs, and shaded bands denote one standard deviation.
(a) Smaller PDPTW instances. (b) Larger PDPTW instances.}
\label{fig:optimality_gap_subplots}
\end{figure*}

Fig.~\ref{fig:optimality_gap_subplots} reports the final relative gap to the
seed-best solution as a function of the number of requests
\(|\mathcal{R}|\). For the smaller instances in
Fig.~\ref{fig:optimality_gap_subplots}(a), the best-performing method depends
on the instance size. Bandit-guided ALNS performs best at
\(|\mathcal{R}|=15\), whereas DQN-guided Hybrid ALNS performs best at
\(|\mathcal{R}|=20\). At \(|\mathcal{R}|=20\), the DQN-guided controller
achieves a mean final gap of about \(0.20\), compared with roughly \(0.29\) for
Bandit-guided ALNS and about \(0.38\)--\(0.39\) for Classical ALNS and
Regret-2. Thus, the learned controller is not uniformly best across all small
instances, but it becomes more effective as the instance size increases within
this group.

For the larger instances in Fig.~\ref{fig:optimality_gap_subplots}(b), the
ranking is again regime-dependent. DQN-guided Hybrid ALNS achieves the lowest
mean gap at \(|\mathcal{R}|=35\), approximately \(0.062\), while
Bandit-guided ALNS performs best at \(|\mathcal{R}|=50\), with a gap of about
\(0.034\). Classical ALNS is also highly competitive at
\(|\mathcal{R}|=50\), outperforming the DQN-guided policy in that setting.

Overall, the online results show that the DQN-guided controller can improve the
mean ALNS search trajectory and final solution quality in several regimes.
However, the advantage is context-dependent rather than universal. This is
consistent with the offline action-space analysis, which showed that quantum
repair actions are admissible only in a subset of reduced repair contexts and
that their benefit depends on the structure and reliability profile of the
local repair state.

\subsection{Fixed-Budget Comparison Across PDPTW Settings}

\begin{figure*}[!h]
\centering
\includegraphics[width=0.8\textwidth]{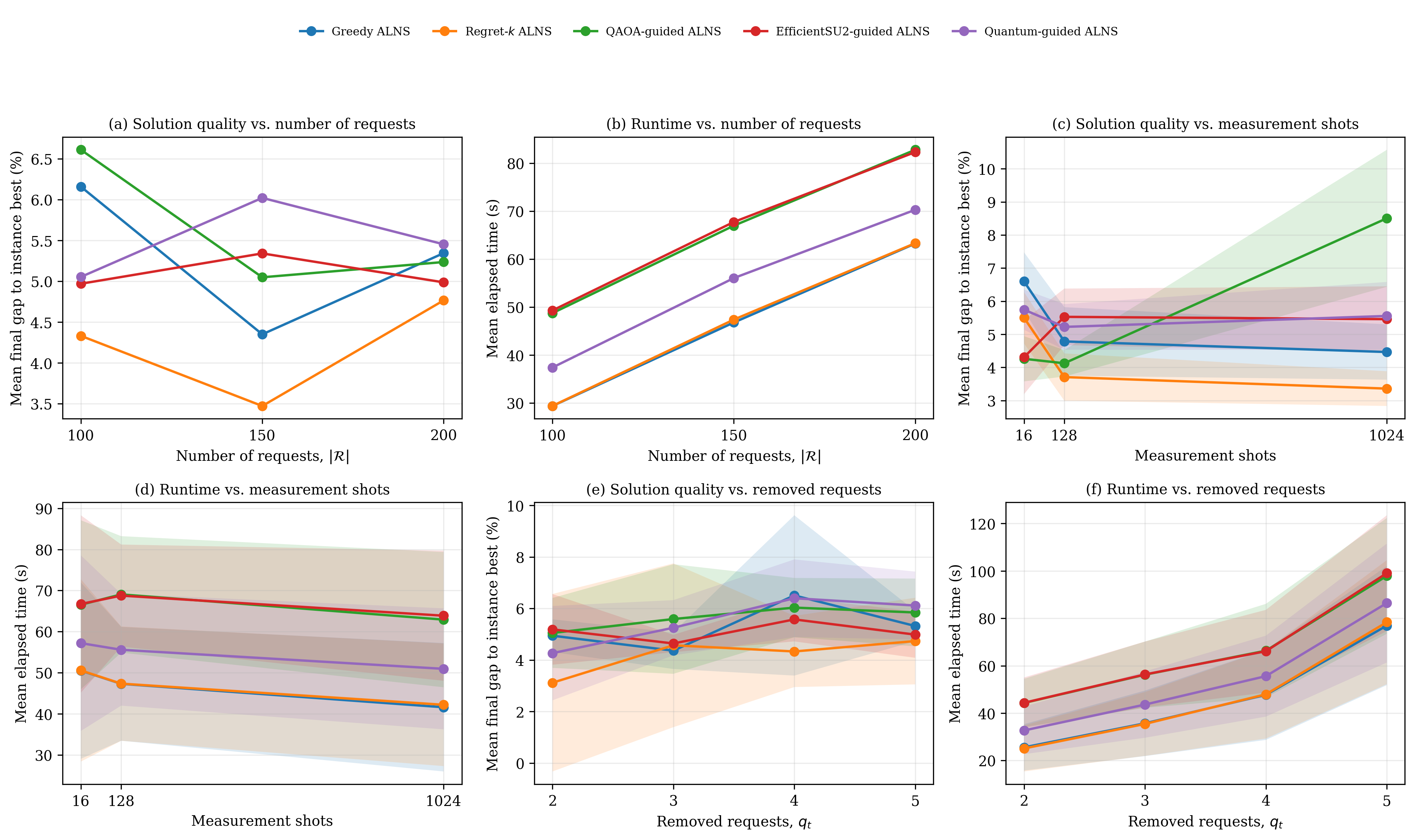}
\caption{Comparison of repair policies under
\((\mathrm{tw\_tightness},\mathrm{capacity\_slack})=(0.15,0.15)\). Curves
report the mean over instance-level averages, and shaded bands show one
standard deviation across instances. (a) Mean final gap to the best objective
found for each instance versus the number of requests \(|\mathcal{R}|\).
(b) Mean elapsed time versus \(|\mathcal{R}|\). (c) Mean final gap versus the
number of measurement shots. (d) Mean elapsed time versus the number of
measurement shots. (e) Mean final gap versus destroy size \(q_t\).
(f) Mean elapsed time versus \(q_t\). Lower values are better in all panels.}
\label{fig:pdptw_setting_015}
\end{figure*}

\begin{figure*}[!h]
\centering
\includegraphics[width=0.8\textwidth]{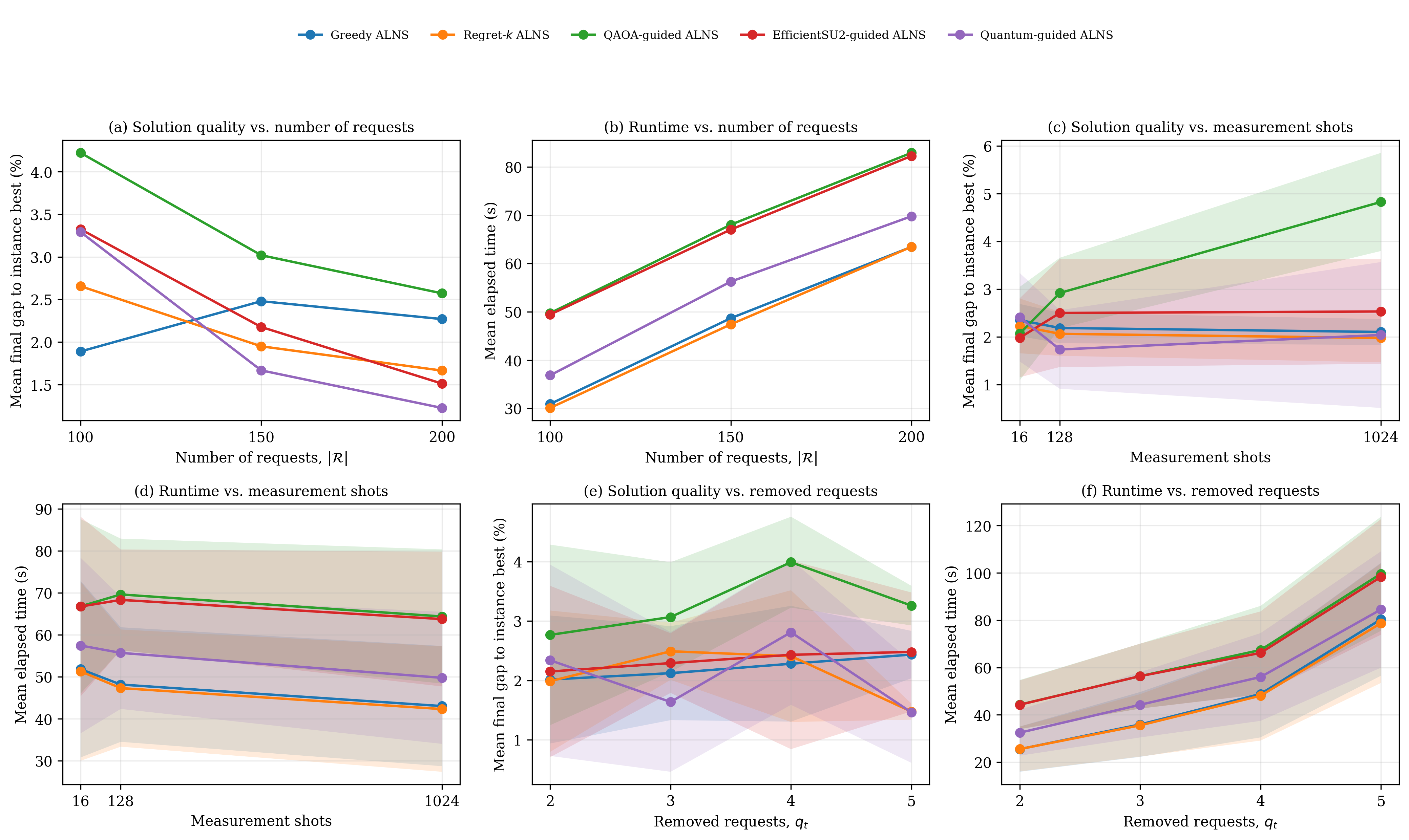}
\caption{Comparison of repair policies under
\((\mathrm{tw\_tightness},\mathrm{capacity\_slack})=(0.85,0.85)\). Curves
report the mean over instance-level averages, and shaded bands show one
standard deviation across instances. (a) Mean final gap to the best objective
found for each instance versus the number of requests \(|\mathcal{R}|\).
(b) Mean elapsed time versus \(|\mathcal{R}|\). (c) Mean final gap versus the
number of measurement shots. (d) Mean elapsed time versus the number of
measurement shots. (e) Mean final gap versus destroy size \(q_t\).
(f) Mean elapsed time versus \(q_t\). Lower values are better in all panels.}
\label{fig:pdptw_setting_085}
\end{figure*}

Figs.~\ref{fig:pdptw_setting_015} and \ref{fig:pdptw_setting_085} compare
repair policies under matched ALNS iteration and repair budgets. Each curve
reports the mean over instance-level averages, and the shaded region denotes
one standard deviation across instances. The plots summarize how final solution
quality and elapsed time change with the number of requests, measurement-shot
budget, and destroy size \(q_t\).

Two PDPTW parameter settings are evaluated. The first setting,
\((\mathrm{tw\_tightness},\mathrm{capacity\_slack})=(0.15,0.15)\), combines
looser time windows with tighter vehicle-capacity conditions. The second
setting, \((0.85,0.85)\), combines tighter time windows with looser
vehicle-capacity conditions. These two settings are therefore not interpreted
as simply ``easy'' or ``hard'' cases; instead, they represent different
constraint regimes.

In the \((0.15,0.15)\) setting, shown in
Fig.~\ref{fig:pdptw_setting_015}, the classical baselines remain strongest
overall. Regret-\(k\) ALNS achieves the lowest mean final gap across most
request-size, shot-budget, and destroy-size panels, while Greedy ALNS remains
competitive and is consistently among the fastest methods. The circuit-only
variants, QAOA-guided ALNS and EfficientSU2-guided ALNS, incur higher runtime
without a corresponding improvement in mean final gap. The selective hybrid
policy reduces part of this circuit-sampling overhead, but its solution quality
is not competitive in this setting. For example, at
\(|\mathcal{R}|=200\), Regret-\(k\) ALNS attains a mean final gap of about
\(4.8\%\), whereas the quantum-guided policy remains above \(5.4\%\). Thus, in
this capacity-constrained regime, selective quantum repair reduces some runtime
cost relative to circuit-only repair but does not improve final solution
quality over the best classical baseline.

In the \((0.85,0.85)\) setting, shown in
Fig.~\ref{fig:pdptw_setting_085}, the ranking changes. The quantum-guided
policy is not best at every operating point, but it becomes the strongest
method as the number of requests increases, achieving the lowest mean final gap
at \(|\mathcal{R}|=150\) and \(|\mathcal{R}|=200\). At
\(|\mathcal{R}|=200\), its mean final gap is about \(1.24\%\), compared with
about \(1.67\%\) for Regret-\(k\) ALNS and \(2.28\%\) for Greedy ALNS. It also
remains faster than the circuit-only QAOA-guided and EfficientSU2-guided
variants, with an elapsed time of about \(70\) s at \(|\mathcal{R}|=200\),
compared with roughly \(82\)--\(83\) s for the circuit-only methods.

The shot-budget and destroy-size panels further show that the hybrid advantage
does not arise simply from using more measurement shots. Instead, the benefit
appears to come from selective use of quantum repair in regimes where the
reduced repair context is large or ambiguous enough for sampling diversity to
be useful, while avoiding the full cost of always invoking a circuit-based
sampler.

Across both settings, three trends are consistent. First, runtime increases
with the number of requests and with destroy size \(q_t\), since both enlarge
the reduced repair space. Second, increasing the number of measurement shots
does not consistently reduce the final gap, especially for QAOA-guided ALNS.
Third, circuit-only repair variants are consistently more expensive than the
classical baselines, while the selective hybrid policy reduces this overhead by
invoking quantum repair only in selected contexts. Overall, the fixed-budget
results support the main conclusion of this study: quantum repair is
context-dependent. It is not globally superior to classical repair, but in
favourable regimes a learned hybrid controller can improve final solution
quality without incurring the full runtime cost of a circuit-only strategy.

\subsection{Successful Quantum Repair Cases on Benchmark Instances}

Table~\ref{tab:quantum_success_cases} reports Li and Lim benchmark instances
for which at least one quantum-enabled repair variant returned a feasible
solution. In selected cases, the returned solution also matched the reported
best-known benchmark distance within numerical tolerance. The strongest cases
are \texttt{lc101}, \texttt{lc201}, and \texttt{lc202}, where QAOA,
EfficientSU2, and the DQN-guided Hybrid ALNS all returned feasible solutions
matching the benchmark distance. These cases demonstrate that circuit-based
local repair can recover benchmark-level solutions when the reduced PDPTW
repair subproblem is sufficiently manageable.

The \texttt{lrc204} instance should be interpreted more cautiously. In this
case, the quantum-enabled variants returned feasible solutions and matched the
best result observed within the experimental grid, but they did not match the
reported best-known benchmark distance. Thus, \texttt{lrc204} provides evidence
of feasible quantum-enabled repair, rather than evidence of a benchmark match.
Overall, the number of successful cases is limited, which is consistent with
the broader results: quantum repair can be effective in selected reduced repair
regimes, but it is not uniformly beneficial across the PDPTW benchmark set.

\begin{table}[!t]
\caption{Benchmark cases with feasible quantum-enabled repair.}
\label{tab:quantum_success_cases}
\centering
\footnotesize
\renewcommand{\arraystretch}{1.08}
\setlength{\tabcolsep}{2.5pt}
\begin{tabular}{p{0.16\columnwidth}p{0.34\columnwidth}p{0.30\columnwidth}p{0.13\columnwidth}}
\toprule
\textbf{Instance} & \textbf{Method(s)} & \textbf{Outcome} & \textbf{Dist.} \\
\midrule
\texttt{lc101}
& QAOA, EfficientSU2, DQN-guided Hybrid ALNS
& Best-known benchmark match
& 828.94 \\
\texttt{lc201}
& QAOA, EfficientSU2, DQN-guided Hybrid ALNS
& Best-known benchmark match
& 591.56 \\
\texttt{lc202}
& QAOA, EfficientSU2, DQN-guided Hybrid ALNS
& Best-known benchmark match
& 591.56 \\
\texttt{lrc204}
& QAOA, EfficientSU2, DQN-guided Hybrid ALNS
& Feasible experimental-best
& 819.76 \\
\bottomrule
\end{tabular}
\end{table}

\section{Conclusion and Future Directions}

This paper presented a noise-aware quantum--classical ALNS framework for the
pickup-and-delivery problem with time windows. Instead of formulating the full
PDPTW as a stand-alone quantum optimization problem, the proposed method embeds
shallow quantum repair samplers inside a classical ALNS backbone. A DQN
controller selects among admissible classical and quantum repair actions using
features that describe the reduced repair context, including entropy,
feasibility, search-space size, search history, and predicted hardware
reliability. Thus, quantum repair is invoked selectively on local repair
subproblems, while the classical ALNS framework preserves global search
robustness and feasibility control.

The results show that the value of quantum repair is regime-dependent. In the
offline dataset, quantum actions are admissible in only \(15.96\%\) of reduced
repair states, and classical repair remains stronger on average under the
current noisy simulation and IBM-backend benchmark model. However, the
fixed-budget benchmark grid shows that quantum-enabled repair can be beneficial
in selected matched regimes. At least one quantum-enabled policy attains a
lower final gap than standard ALNS in 29 of 36 matched settings, with a
best-case gap reduction of \(94.5\%\) and an average reduction of \(50.5\%\)
over the settings in which it is superior. The online ALNS experiments further
show that the learned controller improves mean search behavior in several
regimes, although its advantage is not uniform across all tested instance
sizes.

Together, these findings support the central claim of the paper: the
contribution is not evidence of general quantum advantage, but a learned
decision mechanism for determining when quantum sampling should be used within
a classical heuristic search. The fixed-budget analysis suggests that the
hybrid policy is most useful when the reduced repair context remains rich enough
to contain multiple competing reinsertion patterns, particularly in larger
instances and selected PDPTW parameter regimes. In contrast, when feasibility
filtering makes classical insertion costs highly informative, established
classical repair operators remain difficult to outperform. The hardware
validation results also show a non-negligible mismatch between hardware and
simulation, further motivating the use of an empirical noise-aware model in the
repair controller.

Future work will focus on improving the quality and reliability of the quantum
repair samplers. Promising directions include transferred QAOA parameters,
lightweight offline meta-learning for circuit initialization, adaptive
initialization banks, and noise-aware parameter selection. These strategies aim
to improve sampling quality without introducing the latency of online
variational optimization. Further work should also test the same
regime-selection principle on broader routing and scheduling problems, evaluate
larger and more diverse benchmark sets, and refine how hardware-calibrated
noise models are incorporated into repair-action selection.


\vspace{12pt}
\color{red}

\end{document}